\documentclass{appolb}
\usepackage{epsfig}
\begin{document}
% \eqsec  % uncomment this line to get equations numbered by (sec.num)
\title{Strongly Interacting Matter Matter at Very High Energy Density: 3 Lectures in Zakopane%
\thanks{Presented at the 50'th Crakow School of Theoretical Physics, Zakopane, Poland, June 2010}%
% you can use '\\' to break lines
}
\author{Larry McLerran
\address{Brookhaven National Laboratory and Riken Brookhaven Center\\
Physics Dept. PO Box 5000\\
Upton, NY, 99673, USA}
\and
}

\maketitle
\begin{abstract}
These lectures concern the properties of strongly interacting matter at very high energy density.  I begin with the Color Glass Condensate and the Glasma, matter that controls the earliest times in hadronic collisions.
I then describe the Quark Gluon Plasma, matter produced from the thermalized remnants of the Glasma.
Finally, I describe high density baryonic matter, in particular Quarkyonic matter.  The discussion will be intuitive and based on simple structural aspects of QCD.  There will be some discussion of experimental tests of these ideas.
\end{abstract}

\section{Introduction}

These lectures concern the properties of strongly interacting matter at high energy density.  Such matter occurs in a  number of contexts.  The high density partonic matter that controls the early stages of hadronic collisions at very high energies is largely made of very coherent gluonic fields. In a single hadron, such matter forms the small x part of a wavefunction, a Color Glass Condensate.  After a collision of two hadrons, this matter almost instantaneously is transformed into longitudinal color electric and color magnetic fields.  The ensemble of these fields in their early time evolution is called the Glasma. The decay products of these fields thermalize and form a high temperature gas of quarks and gluons, the Quark Gluon Plasma.  In collisions at lower energy, and perhaps in naturally occurring objects such as neutron stars, there is high baryon density matter at low temperature.  This is Quarkyonic matter.

There is a very well developed literature concerning these various forms of matter.  It is not the purpose of these lectures to provide a comprehensive review.  I will concentrate on motivating and describing such matter from simple intuitive physical pictures and from simple structural aspects of QCD.  I will attempt at various places to relate what is conjectured or understood about such matter to experimental results from accelerator experiments.

\section{Lecture I: The Color Glass Condensate and the Glasma}

The parton distributions of gluons, valence quarks and sea quarks can be measured for some momentum scale less than a resolution scale $Q$ as a function of their fractional momentum $x$ of a high energy hadron.  The lowest value of $x$ accessible for a fixed hadron energy $E$ is typically
$x_{min} \sim \Lambda_{QCD}/E_{hadron}$.  The small x limit is therefore the high energy limit.

\begin{figure}[!htb]
\begin{center}
  \mbox{{\epsfig{figure=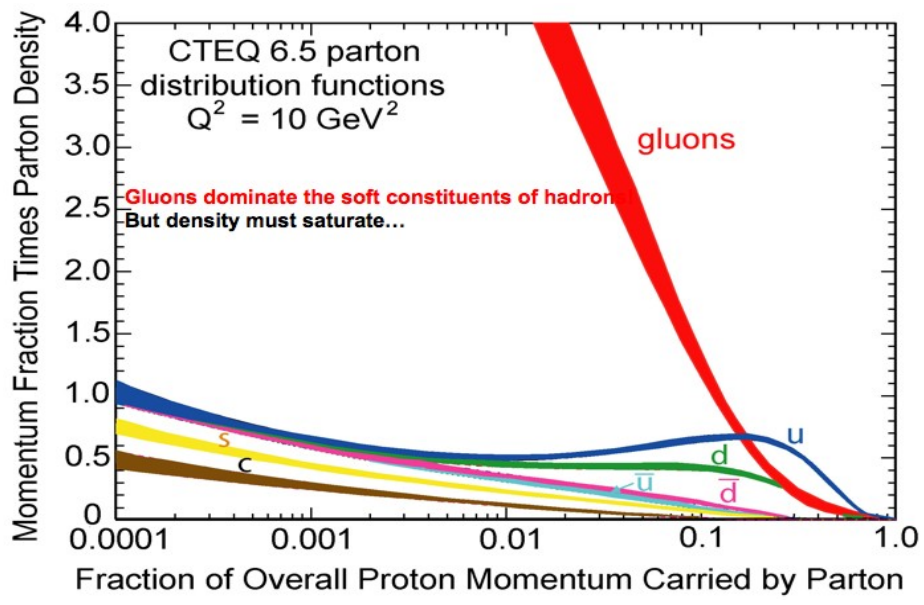, width=0.70\textwidth}}}
   \end{center}
\caption[*]{ \small  The parton distribution as a function of $x$.}
     \label{gluondominance}
\end{figure}

It is remarkable that as $x$ is decreased, as we go to the high energy limit, that the gluon density dominates the constituents of a hadron for $x \le 10^{-1}$.  The various distributions are shown as a function of $x$ in Fig. \ref{gluondominance}.  The gluon density rises like a power of $x$ like $x^{-\delta}$, $\delta \sim .2-.3$
at accessible energies
The area of a hadron grows slowly with energies.  Cross sections grow roughly as $ln^2(1/x)$ for small x.
This means that the rapidly growing gluon distribution results in a high density system of gluons.  At high density, the gluons have small separation and by asymptotic freedom, the intrinsic strength of their interaction must be weak.

A small intrinsic interaction strength does not mean that interactions are weak.  Consider gravity:  The interactions between single protons is very weak, but the force of gravity is long range, and the protons in the earth act coherently, that is always with the same sign.  This results in a large force of gravity.  This can also happen for the gluons inside a hadron, if their interactions are coherent.

\begin{figure}[!htb]
\begin{center}
  \mbox{{\epsfig{figure=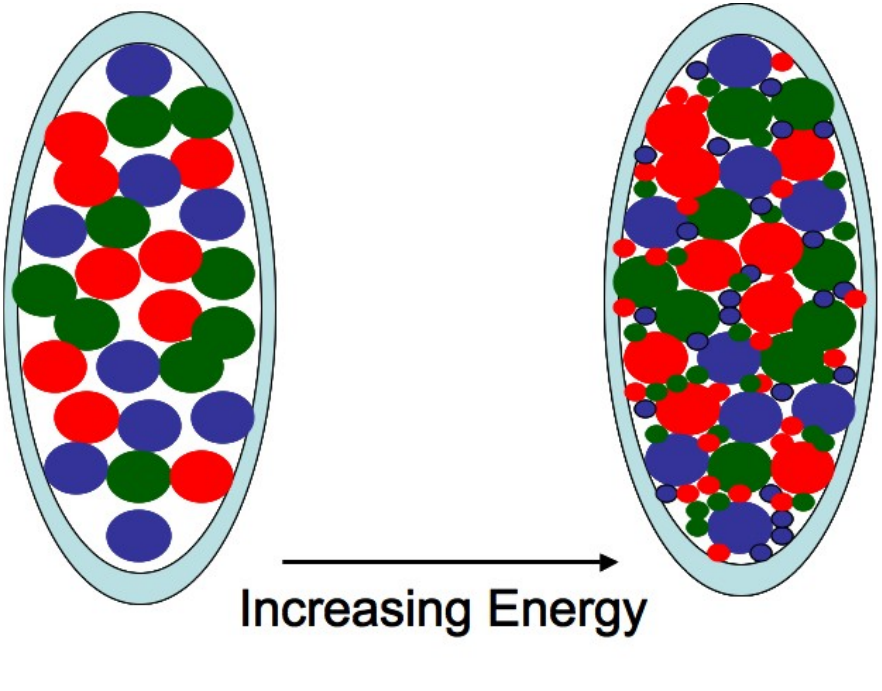, width=0.60\textwidth}}}
   \end{center}
\caption[*]{ \small  Increasing the gluon density in a saturated hadron when going to higher energy.}
     \label{saturation}
\end{figure}

To understand how this might happen, suppose we consider gluons of a fixed size $r_0 \sim 1/p_T$ where
$p_T$ is its transverse momentum.  We assume that at high energy, the gluons have been Lorentz contracted into a thin sheet, so we need only consider the distribution of gluons in the transverse plane.  If
we start with a low density of gluons at some energy, and then evolve to higher energy, the density of gluons increases.  When the density is of order one gluon per size of the gluon, the interaction remains weak because of asymptotic freedom.  When the density is of order $1/\alpha_S$, the coherent interactions are strong, and adding another gluon to the system is resisted by a force of order $1$.  The gluons act as hard spheres. One can add no more gluons to the system of this size.  It is however possible to add in smaller gluons, in the space between the closely packed gluons of size $r_0$.  This is shown in Fig. \ref{saturation}

The physical picture we derive means that below a certain momentum scale, the saturation scale $Q_{sat}$,
the gluon density is saturated and above this scale it is diffuse.  The saturation momentum scale grows with energy and need not itself saturate\cite{Gribov:1984tu}-\cite{McLerran:1993ka}.

The high phase space density of gluons, $dN/dyd^2p_Td^2r_T \sim 1/\alpha_S$ suggests that one can describe the gluons as a classical field.  A phase space density has a quantum mechanical interpretation
as density of occupation of quantum mechanical states.  When the occupation number is large, one is in the classical limit.

One can imagine this high density gluon field generated from higher momentum partons.  We introduce the idea of sources corresponding to high $x$ partons and fields as low $x$ partons.  Because the high $x$ parton sources are fast moving, their evolution in time is Lorentz time dilated.  The gluon field produced by these sources is therefore static and evolves slowly compared to its natural time scale of evolution.  This ultimately means that the different configurations of sources are summed over incoherently, as in a spin glass. 

We call this high energy density configuration of colored fields a Color Glass Condensate.  The word color is because the gluons that make it are colored.  The word condensate is used because the phase space density of gluons is large, and because this density is generated spontaneously.  The word glass is used because the typical time scale of evolution of the classical fields is short compared to the Lorentz time dilated scales associated with the sources of color.

There is an elaborate literature on the Color Glass Condensate and an excellent review is by Iancu and Venugopalan\cite{Iancu:2003xm}.  Evolution of the CGC to small values of x is understood, as well as many relationships between deep inelastic scattering, deep inelastic diffraction and high energy nucleus-nucleus,
proton-nucleus and proton-proton scattering.  The CGC is a universal form of matter in the high energy limit.  The theoretical ideas underlying the CGC are largely unchallenged as a description of the high energy limit of QCD, but the issue of when the approximation appropriate for the high energy limit are valid remains contentious. 

\begin{figure}[!htb]
\begin{center}
  \mbox{{\epsfig{figure=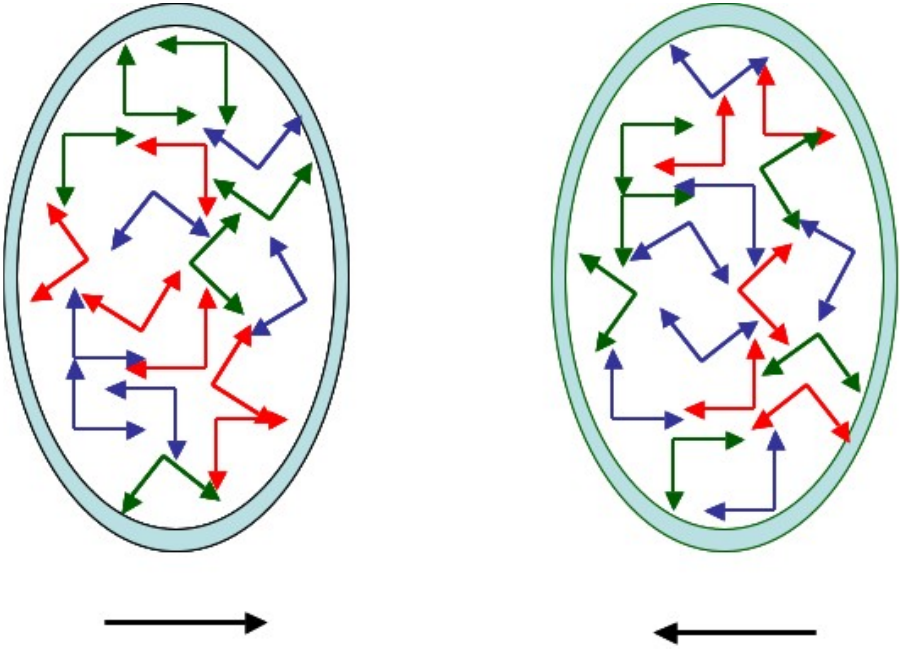, width=0.70\textwidth}}}
   \end{center}
\caption[*]{ \small  The collision of two sheets of CGC.}
     \label{sheets}
\end{figure}

In the description of high energy hadron hadron collisions, we consider the collision of two sheets of CGC as shown in Fig. \ref{sheets}.  The color electric and color magnetic fields of the CGC are visualized as sheets of Lenard-Wiechart potentials.  These are classical gluon fields whose polarization and color are random, with an intensity distribution determined by the underlying theory of the CGC.

Upon collision of these sheets, the sheets become charged with color magnetic and color electric charge distributions of equal magnitude but opposite sign locally in the transverse plane of the sheets\cite{Kovner:1995ja}-\cite{Lappi:2006fp}. In the high
energy limit sources of color electric and color magnetic field must be treated on an equal footing because of the self duality of QCD.  This induced charge density produces longitudinal color electric and color magnetic fields between the two sheets.  These fields are longitudinally boost invariant and therefore have the correct structure to account for Bjorken's initial conditions in heavy ion collisions\cite{Bjorken:1982qr}.  The typical transverse length scale over which the flux tubes vary is $1/Q_{sat}$. The initial density of produced gluons is on dimensional grounds
\begin{equation}
  {1 \over {\pi R^2}} {{dN} \over {dy}} \sim {{Q_{sat}^2} \over {\alpha_S}}
\end{equation}
Because there are both color electric and color magnetic fields, there is a topological charge density of maximal strength induced $FF^D \sim Q_{sat}^2/\alpha_S$  
\begin{figure}[!htb]
\begin{center}
  \mbox{{\epsfig{figure=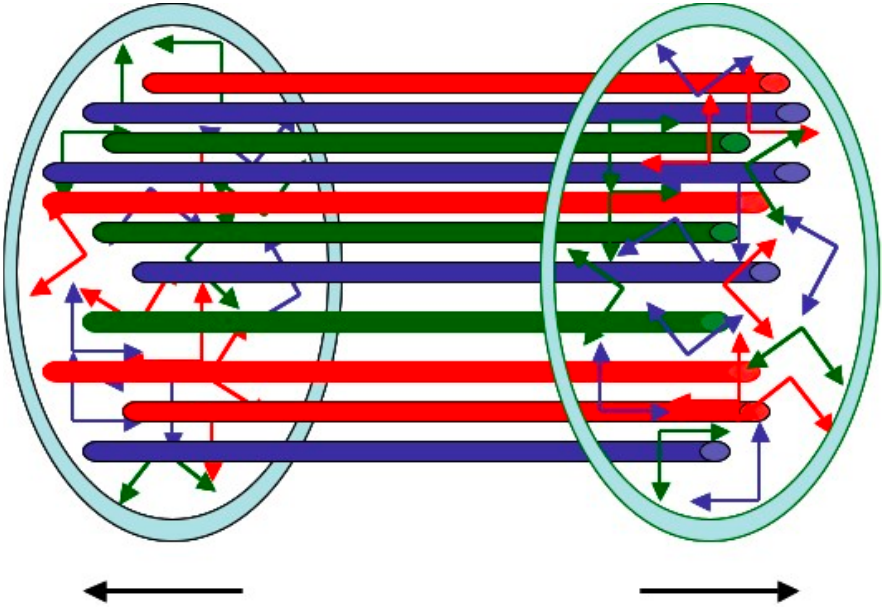, width=0.70\textwidth}}}
   \end{center}
\caption[*]{ \small  Glasma flux tube produces after the collision.}
     \label{glasma}
\end{figure}

The decay of products of the Glasma is what presumably makes a thermalized Quark Gluon Plasma.  It is not clear how this thermalization takes place.  It is quite likely that in the decay of these fields, a turbulent fluid arises, and perhaps this fluid can generate an expansion dynamics similar to that of a thermalized QGP for at least some time\cite{Dusling:2010rm}.

\subsection{The CGC and Electron-Hadron Scattering}

If the only momentum scale that controls high energy scattering is the saturation momentum,
then there will be scaling\cite{Stasto:2000er}.  In particular, the cross section for deep inelastic scattering will be a function
\begin{equation}
\sigma_{\gamma^*p} \sim F(Q^2/Q_{sat}^2)
\end{equation}
rather than a function of $Q^2$ and $x$ independently.  The x dependence of the saturation momentum may be determined empirically as $Q_{sat}^2 \sim Q_0^2/x^\delta$ where $\delta = 0.2-0.3$, which is consistent with analysis of evolution equations\cite{Balitsky:1995ub}-\cite{Mueller:2002zm}.
The scaling relationship can be derived from the classical theory for $Q^2 \le Q_{sat}^2$.  It can further be shown to extend over a much larger range of $Q^2$\cite{Iancu:2002tr}.  For large values of $Q^2$ this scaling is a consequence of the linear evolution equations, but the global structure is determined by the physics of saturation.  Such a simple scaling relationship describes deep inelastic scattering data for $ x \le 10^{-2}$.
\begin{figure}[!htb]
\begin{center}
  \mbox{{\epsfig{figure=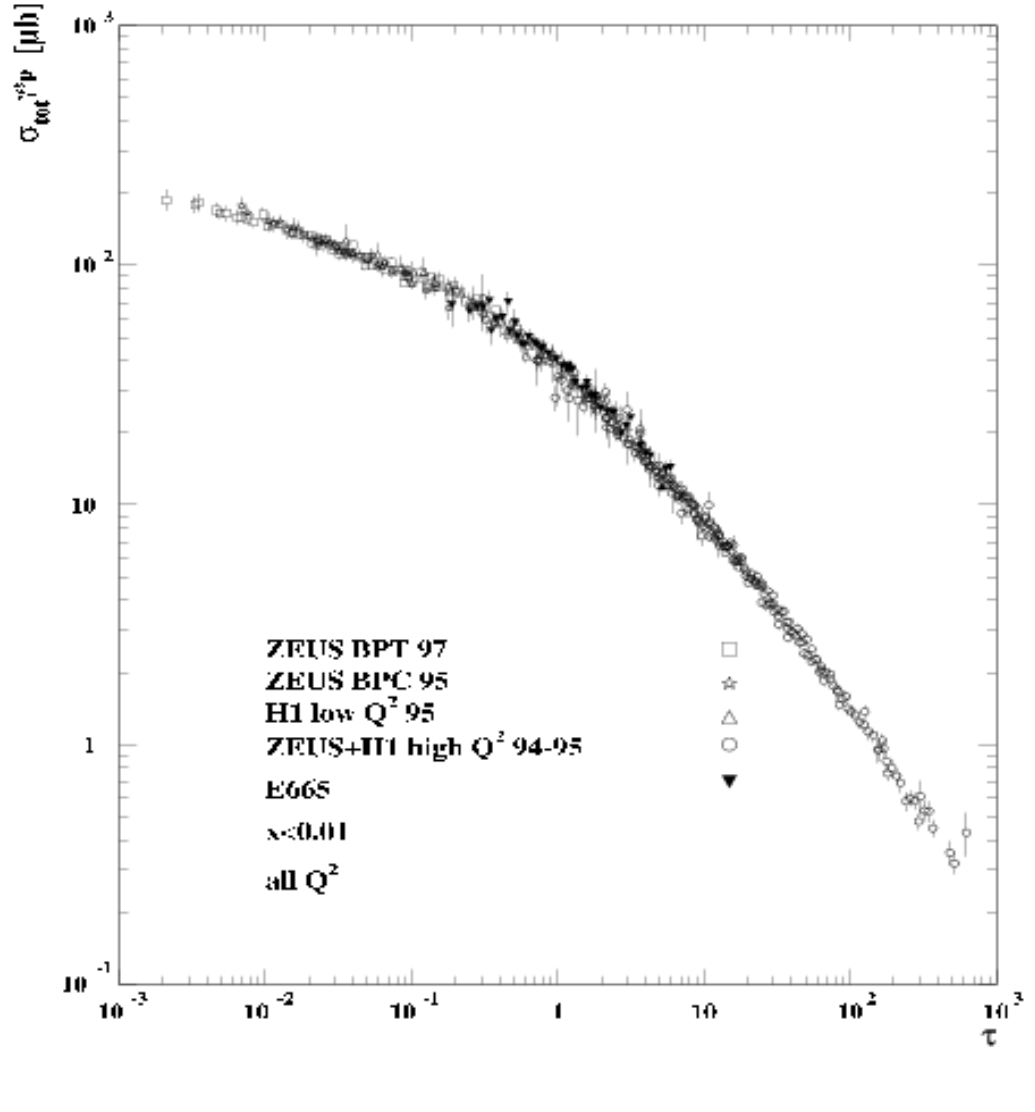, width=0.70\textwidth}}}
   \end{center}
\caption[*]{ \small  The geometric scaling in deep inelastic scattering.}
     \label{geometric}
\end{figure}

\begin{figure}[!htb]
\begin{center}
  \mbox{{\epsfig{figure=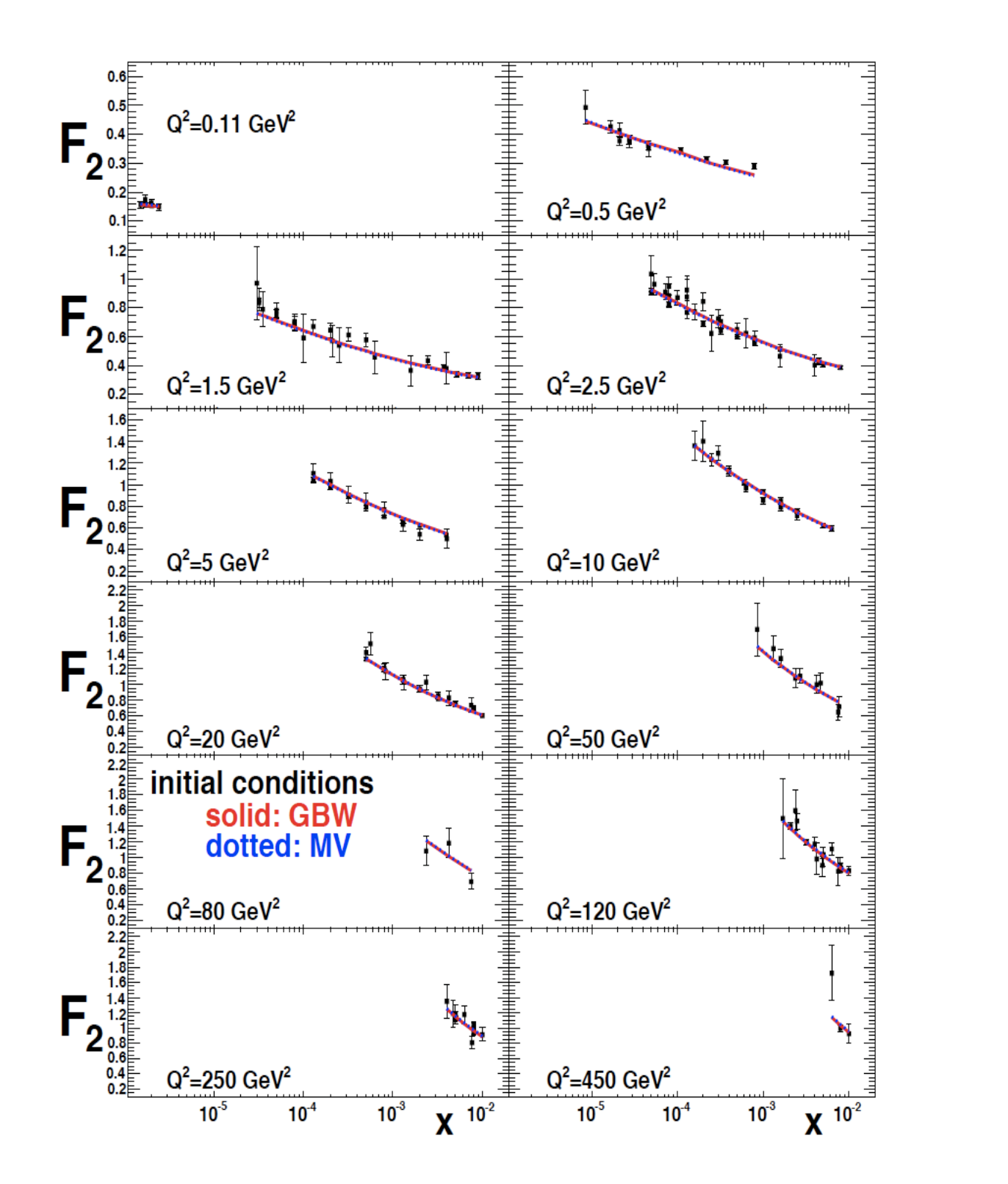, width=0.90\textwidth}}}
   \end{center}
\caption[*]{ \small  The CGC description of F2 data in deep inelastic scattering.}
     \label{f2}
\end{figure}

Using evolution equations for the CGC including the effects of running coupling constant\cite{Albacete:2007yr}-\cite{Balitsky:2008zza}, one can compute
deep inelastic scattering structure functions at small x\cite{Albacete:2009fh}.  This involves very few parameters, and provides comprehensive description of deep inelastic scattering data at $x \le 10^{-2}$.
The description of $F_2$ in deep inelastic scattering is shown in Fig. \ref{f2}.  It should be noted that in
the CGC description of deep inelastic scattering, the gluon distribution function is the Fock space distribution of gluons inside a hadron.  It can never become negative.  In the description of the $F_2$ data,
the gluon distribution function is not becoming small at small $Q^2$ as is the case in some linear evolution fits.  This is intuitively reasonable since we have no reason to expect that the Fock space distribtuion of gluons in a hadron should become small at small $Q^2$.

The Color Glass Condensate description may also be applied to diffractive deep inelastic scattering,
and with the same parameters that describe deep inelastic scattering does an excellent job of describing
the data.  In addition, there are measurements of the longitudinal structure function, a quantity directly proportional to the gluon density.  Conventional descriptions that use linear DGLAP evolution equations are somewhat challenged by this data, but the CGC description naturally fits the data.  

To summarize, the CGC description of deep inelastic scattering at small x naturally describes $F_2$, $F_L$
and diffractive data. It is a successful phenomenology  Why is the CGC therefore not accepted as the standard description? The problem is that the linear evolution DGLAP descriptions describe $F_2$ adequately, except in the region where the perturbative computations most probably breaks down.  They do not do a very good job on the low $Q^2$ $F_L$ data, but this is where there is a fair uncertainty in the data.
The diffractive data is naturally described in the CGC framework, but there are other successful models.
Ultimately, there is no consensus within the deep inelastic scattering community that the CGC is needed in order to describe the data.

\subsection{The CGC and Heavy Ion Collisions}

\subsubsection{Multiplcities in RHIC Nulcear Collsions}

\begin{figure}[!htb]
\begin{center}
  \mbox{{\epsfig{figure=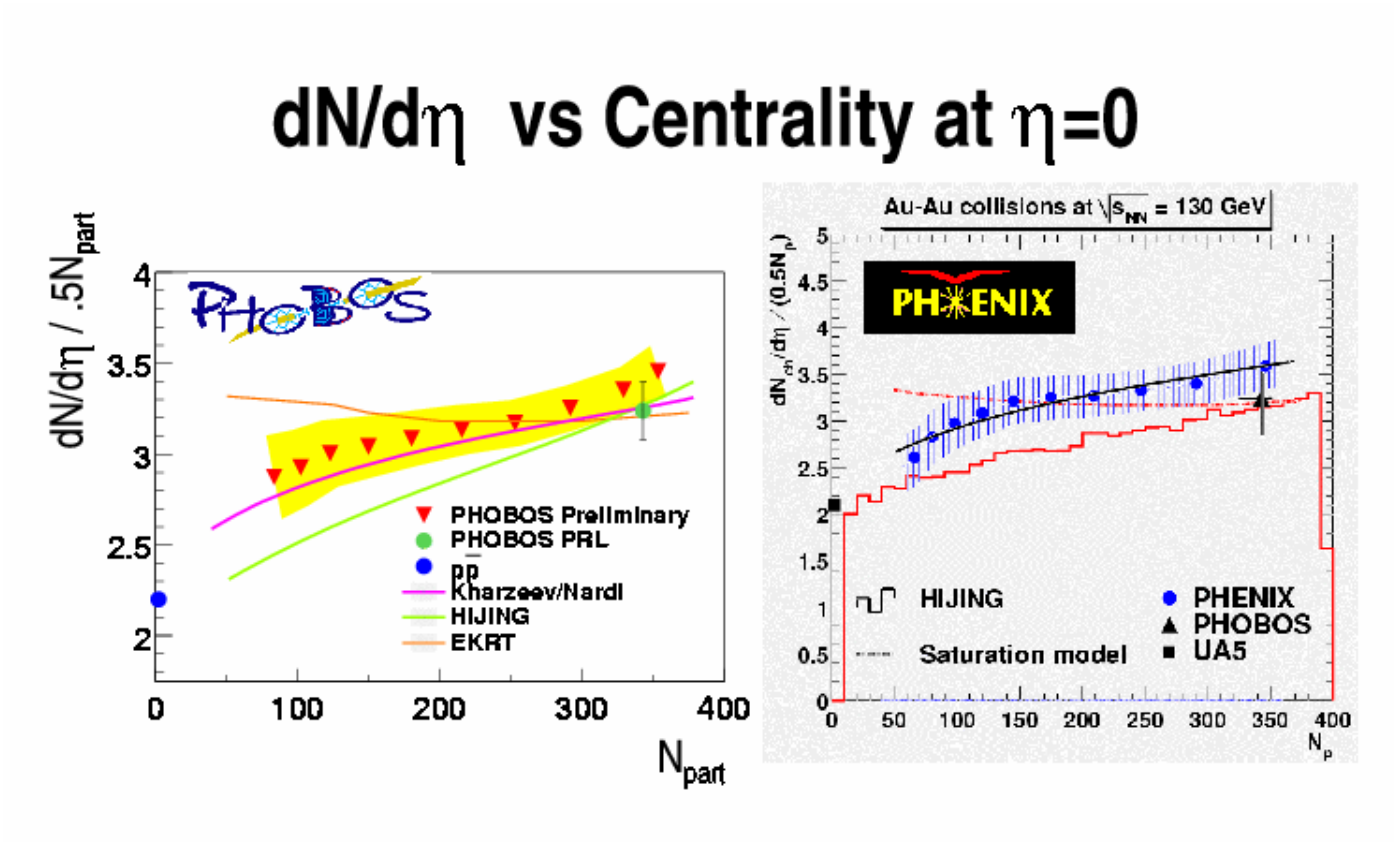, width=0.90\textwidth}}}
   \end{center}
\caption[*]{ \small  The multiplicity as a function of the number of nucleon participants in heavy ion collisions.}
     \label{aa}
\end{figure}

One of the early successes of the CGC was the description of multiplicity distributions in deep inelastic scattering\cite{Kharzeev:2000ph}-\cite{Kharzeev:2001yq}.  Recall that the phase space distribution of gluons up to the saturation momentum is of order $Q_{sat}^2/\alpha_S(Q_{sat})$.  We will assume that the distribution of
initially produced gluons is proportional to this distribution of gluons in the hadron wavefunctions of the colliding nuclei and further,that the multiplicity of produced gluon is proportional to the final
state distribution of pions.  We get
\begin{equation}
{1 \over \sigma}~  {{dN} \over {dy}} \sim  {1 \over{ \alpha_S(Q_{sat})}}Q_{sat}^2 \sim A^{1/3} x^{-\delta}
\end{equation}  
Here $\sigma$ is the area of overlap of the two nuclei in the collision and A the number of nucleons that participate in the collision.  $\sigma ~Q_{sat}^2 \sim A$
at low energies assumes no shadowing of nucleon parton distributions and is consistent with
information concerning deep inelastic scattering on nuclear targets.  In the collisions of nuclei one can directly measure the number of nucleonic participants in the collisions, a number that varies with the centrality of the collision. One can then compare the central region multiplicity with the number of participants so determined.  Such a comparison is shown in\cite{Adcox:2004mh}-\cite{Back:2004je} Fig. \ref{aa} .
The saturation description of Kharzeev and Nardi provides a good description of the centrality dependence of the collisions\cite{Kharzeev:2000ph}.  It also does well with the energy dependence.  Refinements of this description can provide a good description of the rapidity distribution of produced particles\cite{Kharzeev:2001yq}.

\subsubsection{Limiting Fragmentation}

\begin{figure}[!htb]
\begin{center}
  \mbox{{\epsfig{figure=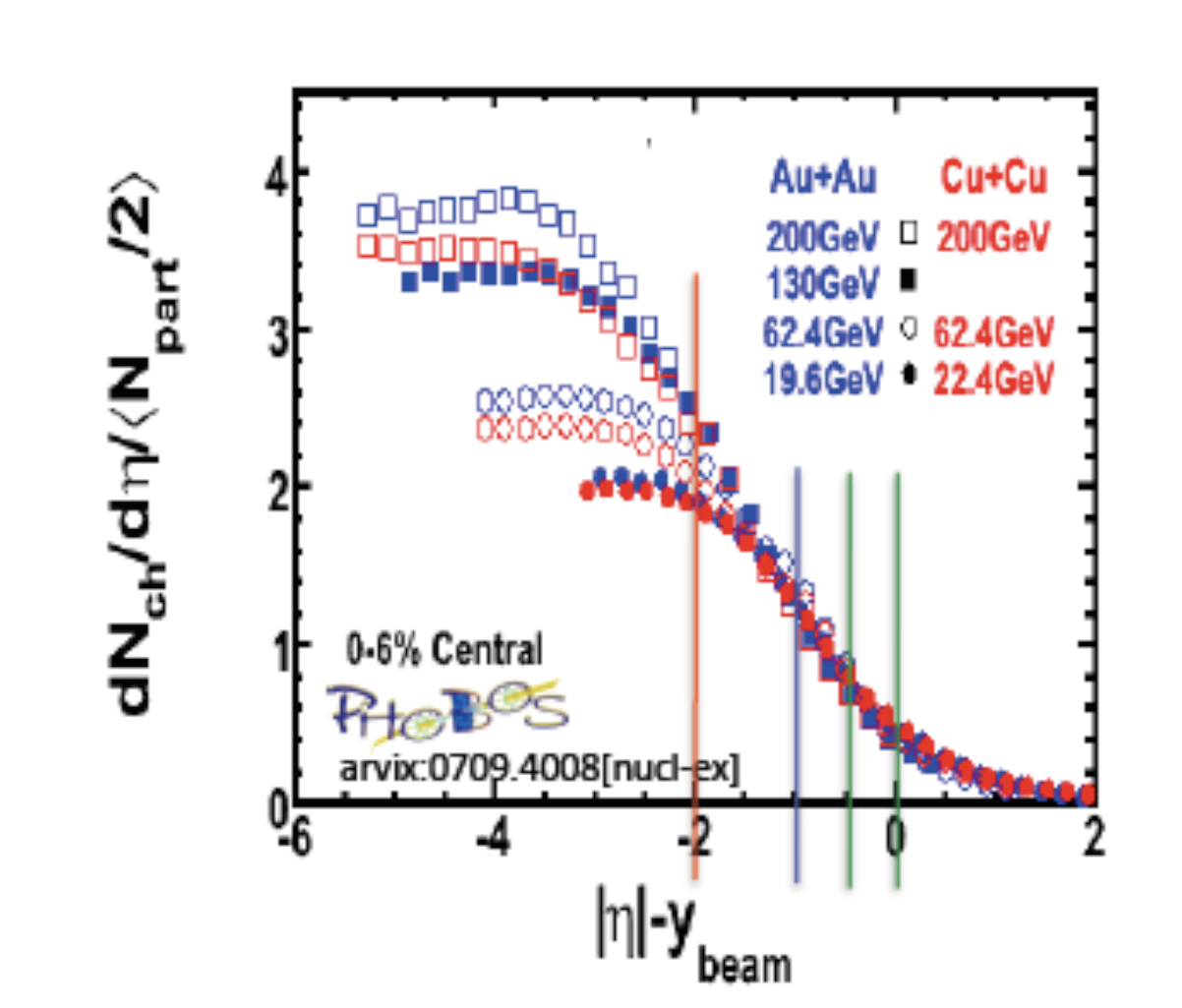, width=0.90\textwidth}}}
   \end{center}
\caption[*]{ \small  Limiting fragmentation in RHIC nuclear  collisions.}
     \label{limfrag}
\end{figure}

A general feature of high energy hadronic scattering is limiting fragmentation.  If one measures the distribution of particles as a function of rapidity up to some fixed rapidity from the rapidity of one of the colliding particles, then the distribution is independent of collision energy.  The region over which this scaling occurs
increases as the energy of the colliding particles increases.  Such scaling is shown in Fig. \ref{limfrag}.
Such limiting fragmentation is natural in the CGC approach.  For example in Fig. \ref{limfrag},
we see that the region of limiting fragmentation increases as beam energy increases\cite{Back:2004je}.  If we think of the region where there is limiting fragmentation as sources for fields at small more central rapidities, then we see that going to higher energies corresponds to treating a larger region as sources.  In a renormalization group language, this simply means that one is integrating out fluctuations at less central rapidities, to generate an effective theory for the particles at more central rapidity.  A quantitative description of limiting fragmentation within the theory of the  CGC is found in Ref.   \cite{Gelis:2006tb}.

\subsubsection{Single Particle Distributions in dAu Collisions}

\begin{figure}[!htb]
\begin{center}
  \mbox{{\epsfig{figure=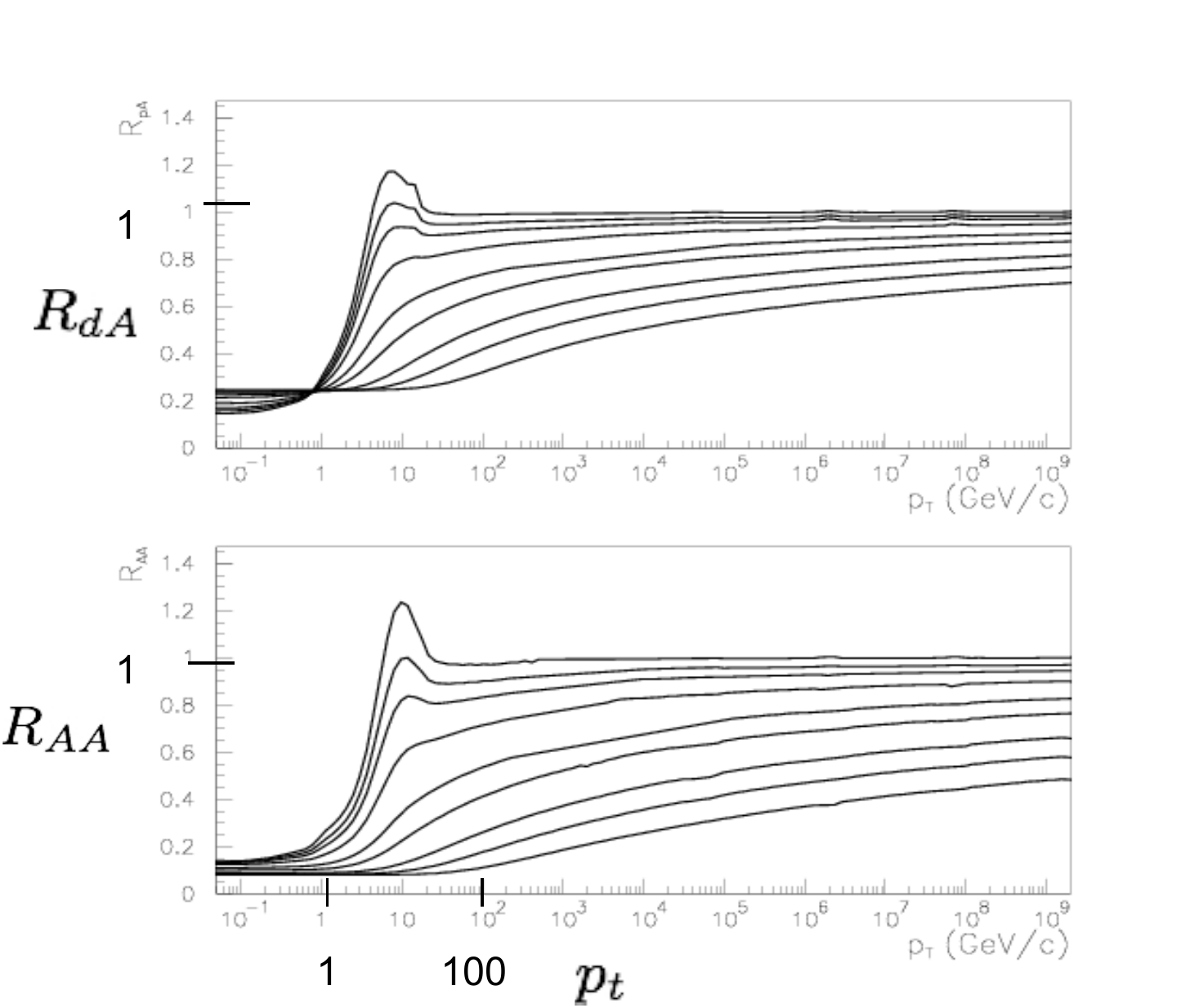, width=0.90\textwidth}}}
   \end{center}
\caption[*]{ \small  The ratio of particles emitted in dA  and AA collisions to that in proton due to CGC effects.}
     \label{dA}
\end{figure}

Some of the early predictions of the CGC were generic features of the single particle inclusive distributions
seen in hadron-nucleus collisions.  There are two competing effects.  The first is multiple scattering of a hadron as it traverses a nucleus.  This effect is included n the CGC gluon distributions as an enhancment
of the gluon distribution for  $p_T$ at transverse momentum of the order of the saturation momentum,
with a corresponding depletion at smaller momentum.  There is little effect at high $p_T$.  The other effect is that in the evolution of the gluon distribution to small $x$, the saturation momentum acts as a cutoff in the
bremstrahlung like integrals that generate such small x gluons.  Nuclei have a larger saturation momentum
than do hadrons, so the small x gluon distribution for nuclei will be suppressed relative to that for hadrons.  Put another way, this effect will generate a suppression for more central collisions.  The sum of these effects is shown in Fig. \ref{dA}\cite{Baier:2003hr}-\cite{Iancu:2004bx}. 
The different curves correspond to different rapidities of the produced particle,
beginning with the top curve being near the fragmentation region of the nucleus.  As one evolve further in rapidity, the enhancement at intermediate transverse momentum disappears and is replaced by a smooth curve with an overall suppression of produced particles.

The pattern of suppression suggested by the Color Glass Condensate was first seen in dAu collisions
in the Brahms collaboration\cite{Arsene:2004fa}, and later confirmed by the other experiments \cite{Adcox:2004mh},\cite{Back:2004je},\cite{Adams:2005dq}.  The Brahms experiments demonstrated
that in the nuclear target fragmentation region that at intermediate $p_T$ there was en enhancement in $R_{dA}$
as a function of centrality, but in the deuteron fragmentation region, there was a depletion as a function of centrality.  The CGC provided the only model that predicted such an effect, and it remains the only 
theory that can quantitatively explain the suppression seen in the deuteron fragmentation region.

\subsubsection{Heavy Quark and $J\Psi$ Production}

If the saturation momentum is small compared to a quark mass, it can be treated as very heavy.  It should have perturbative incoherent production cross sections.  If the saturation momentum is large compared to a quark
mass, the quarks should be thought of as light mass.  Cross sections for production should be coherent,
and for example in $pA$ collisions, scale as $A^{2/3}$.   In the deuteron fragmentation region of dAu collisions we would expect suppression of heavy quark and charmonium cross sections relative to the nuclear fragmentation region. 
\begin{figure}[!htb]
\begin{center}
  \mbox{{\epsfig{figure=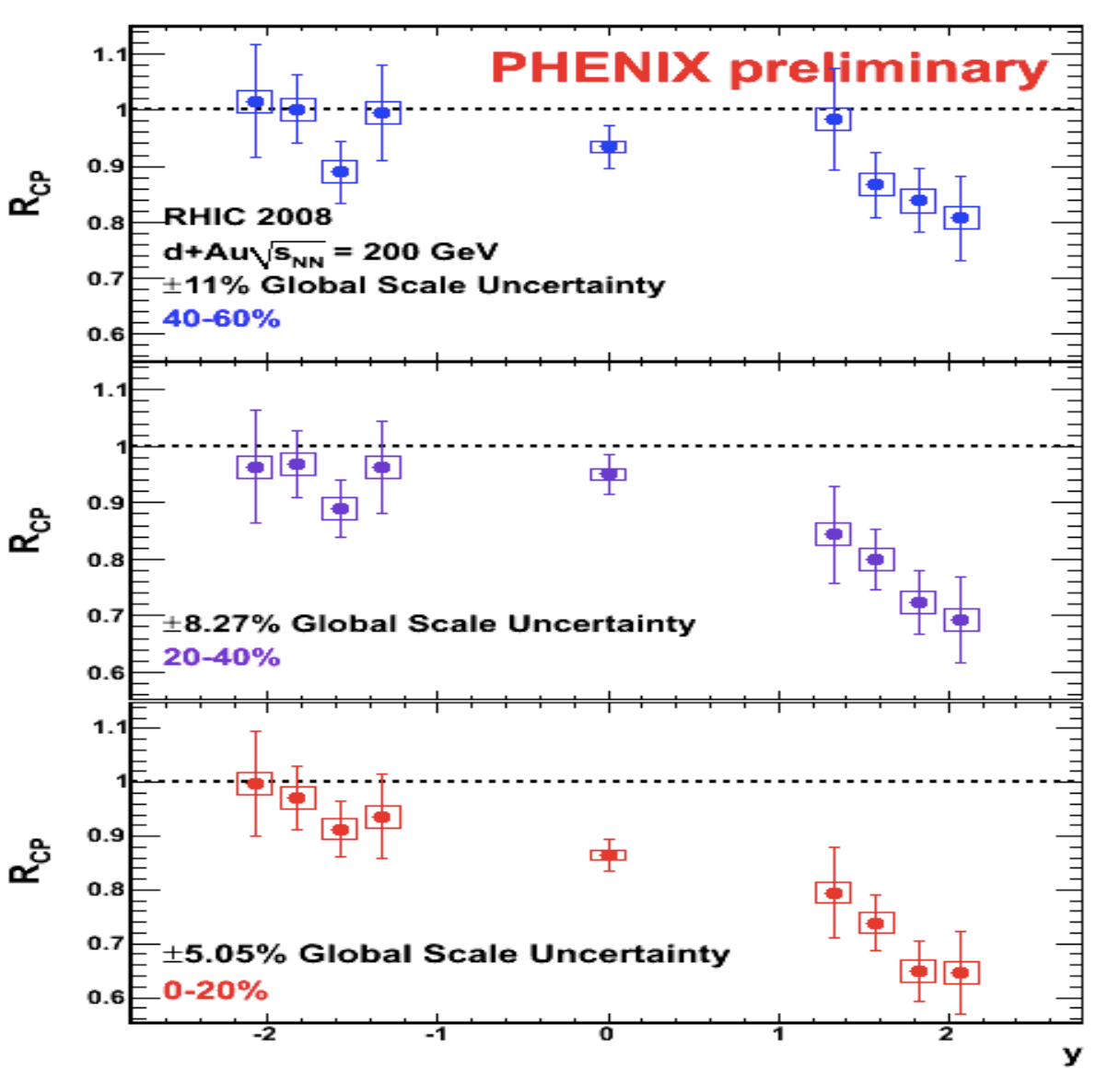, width=0.90\textwidth}}}
   \end{center}
\caption[*]{ \small  The $J/\Psi$ production cross section as a function of centrality and rapidity.}
     \label{jpsi}
\end{figure}
 In Fig. \ref{jpsi}, the ratio of central to peripheral cross sections for $J/\Psi$ production is shown as a function of centrality and rapidity.  Note the strong suppression in the forward region for central collisions, as expected from the CGC.  Precise computations are difficult for the charm quark since its mass is close to the saturation momentum.  Such computations are in agreement with the data at forward rapidity\cite{Kharzeev:2008nw}-\cite{Kopeliovich:2010nw}.
 
 \subsubsection{Two Particle Correlations}
 
\begin{figure}[!htb]
\begin{center}
  \mbox{{\epsfig{figure=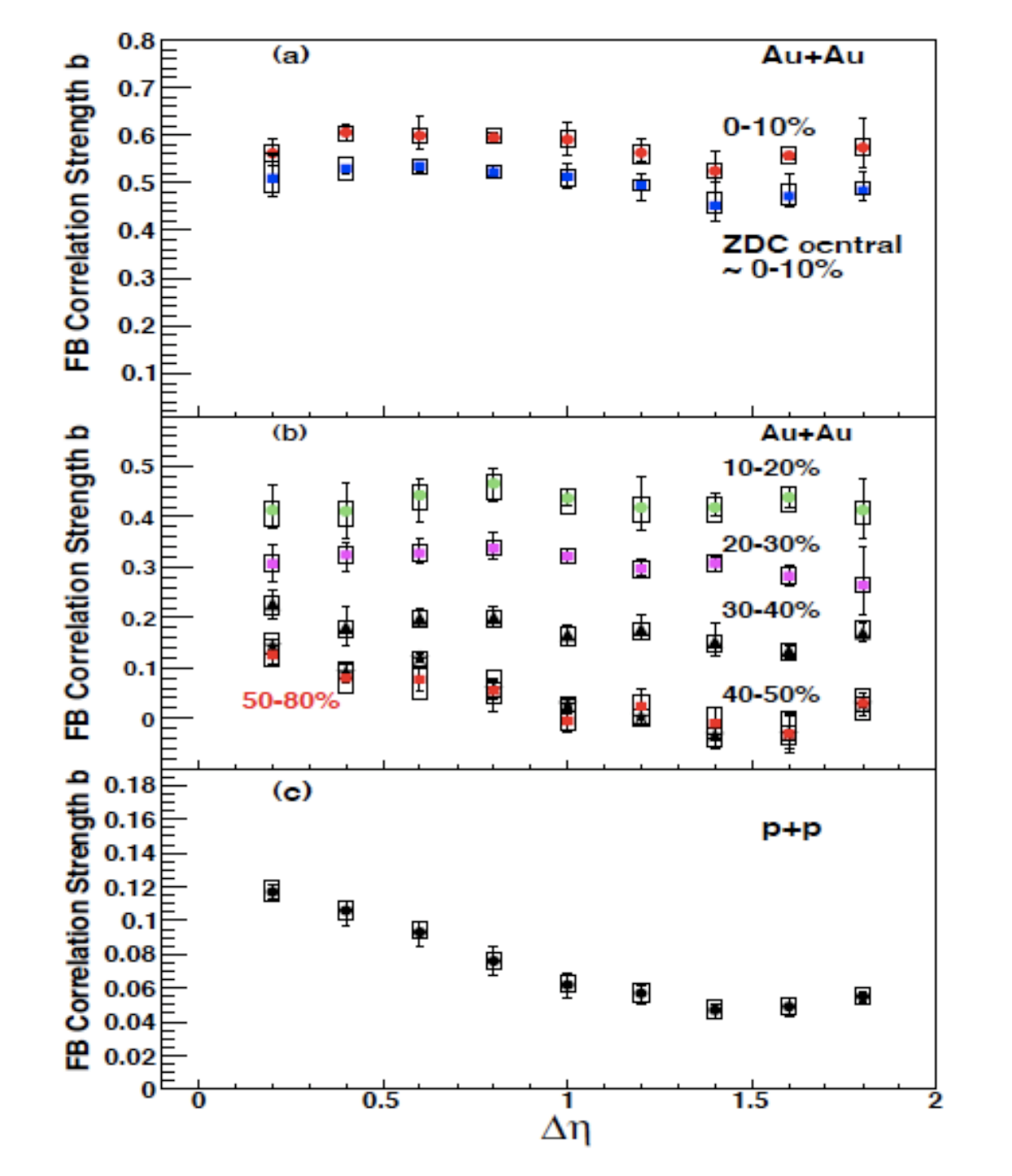, width=0.90\textwidth}}}
   \end{center}
\caption[*]{ \small  The strength of forward backward correlations as a function of rapidity and centrality.}
     \label{fb}
\end{figure}

 The Glasma flux tubes induced by the collision of two hadrons will generate long range correlations in rapidity.  In heavy ion collisions, this may be seen in forward backward correlations, as measured in STAR.  The correlation increases in strength with higher energy collisions or more central collisions.  This is expected in the CGC-Glasma description because for more central collisions the saturation momentum
 is bigger, so that the system is more correlated.  (The coupling becoming weaker means the system is more
 classical, and therefore the leading order contribution associated with Glasma flux tubes becomes
 relatively more important.)  Such forward-backward correlations are shown in Fig. \ref{fb} as a function of rapidity and centrality\cite{:2009dqa}-\cite{Lappi:2009vb}.  The value of the correlation coefficient b
 can be shown to be bounded $b \le 1/2$.

Such two particle correlations in the Glasma can generate ridge like structures seen in two particle correlation experiments in azimuthal angle and rapidity\cite{Shuryak:2007fu}-\cite{Dumitru:2008wn}.  The long range rapidity correlation is intrinsic to the Glasma.  The angular correlation might be generated by flow effects at later times in the collision, by opacity and trigger bias effects, or by
intrinsic angular correlations associated with the decay of Glasma flux tubes\cite{Gavin:2008ev}-\cite{Dumitru:2010iy}.

\subsubsection{The Negative Binomial Distribution}

The decay of a single Glasma flux tube generates a negative binomial distribution of produced particles\cite{Gelis:2009wh}.
A sum of negative binomial distributions is again a negative binomial distribution.  Such oa form of the distribution describes the RHIC data well.  It is difficult with the heavy ion data to isolate those effects due to an intrinsic negative binomial distribution and those due to impact parameter.  It is possible to isolate the effects of impact parameter, but it demands a high statistics study.  

\subsubsection{Two Particle Azimuthal Angular Correlations in dA Collisions}

\begin{figure}[!htb]
\begin{center}
  \mbox{{\epsfig{figure=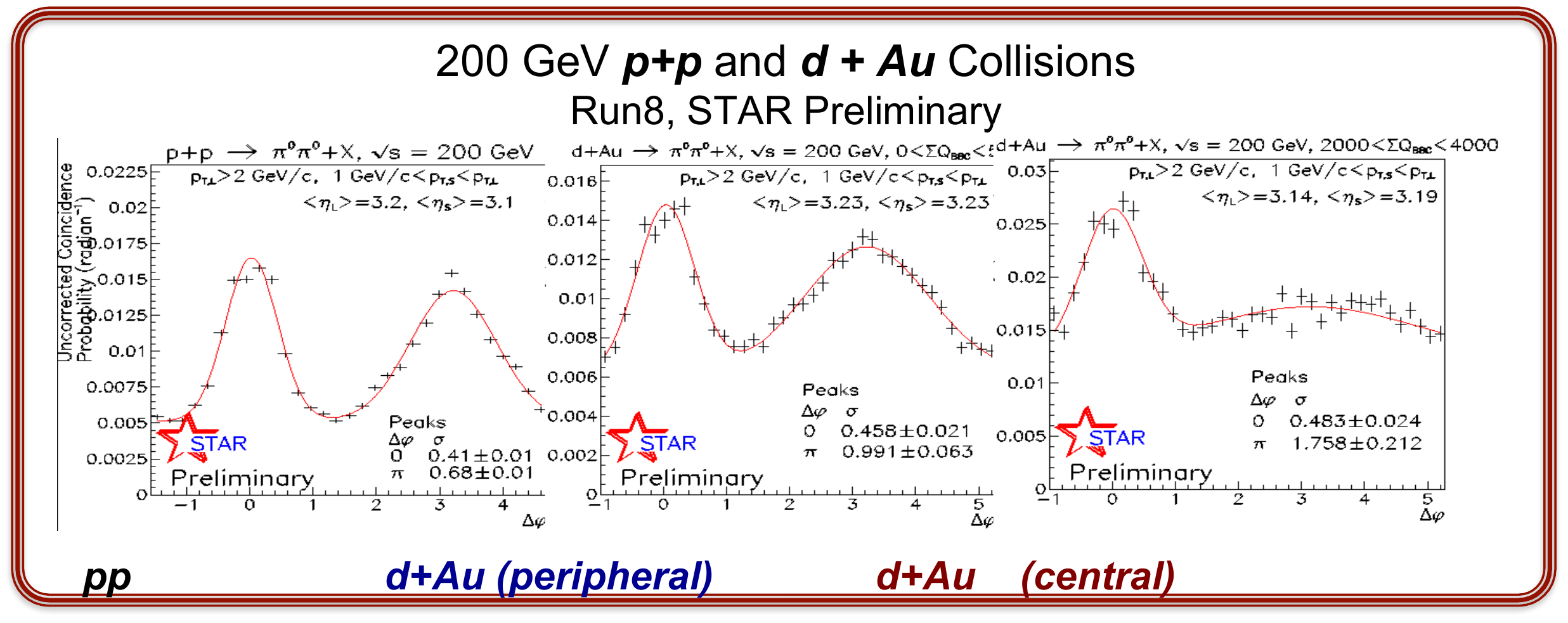, width=0.90\textwidth}}}
   \end{center}
\caption[*]{ \small Forward rapidity, forward backward angular correlations in dAu collisions
as a function of centrality.}
     \label{dafb}
\end{figure}

The CGC will de-correlate forward-backward angular correlations when the the transverse momentum of produced particles is of order the saturation momentum\cite{Kharzeev:2004bw}-\cite{Albacete:2010rh}.
This is because near the produced particles get momentum from the CGC and
therefore are not back-to-back correlated.  In dAu collisions such an effect will be largest at forward rapidities near the fragmentation region of the deuteron, since this corresponds to the smallest values of x for the nuclear target.  This kinematic region is least affected by multiple scattering on the nucleus.  This effect has been seen by the STAR and PHENIX collaborations\cite{Braidot:2010ig}-\cite{Meredith:2009fp}.  There is a good quantitative description by Tuchun and by Albacete and Marquet,\cite{Tuchin:2009nf}-\cite{Albacete:2010rh}  as shown in the figure \ref{dafb}

\subsection{Concluding Comments on the CGC and the Glasma}

There is now a wide variety or experimental data largely consistent with the CGC and Glasma based description.  There is a well developed theoretical framework that provides a robust phenomenology
of both electro-hadron scattering and hadron scattering,  There are new areas that are developing that I have not had time to discuss.  One is the possibility to see effects of topological charge change in heavy ion collisions, the Chiral Magnetic Effect\cite{Kharzeev:2007jp}.  Another area is pp collisions at the LHC, where some work concerning recent experimental data was developed at this school\cite{McLerran:2010ex}.

\section{Lecture II: Matter at High Temperature: The Quark Gluon Plasma}

\section{Matter at Finite Temperature}

\subsection{Introdcution}

In this lecture I will describe the properties of matter at high temperature.  The discussion here will be theoretical.  There is a wide literature on the phenomenology of the Quark Gluon Plasma and its possible
description of heavy ion collisions at RHIC energies.  The interested reader is referred to that literature.
I will here develop the ideas of decofinement, chiral symmetry restoration based in part on a simple description using the large number of colors limit of QCD.

\subsection{Confinement}

The partition function is
\begin{equation}
 Z = Tr~e^{-\beta H + \beta \mu_B N_B}
\end{equation}
where the temperature is $T = 1/\beta$ and $N_B$ is the baryon number and $\mu_B$ is the baryon number chemical potential.  Operator expectation values are
\begin{equation}
 <O> = {{Tr~ O ~e^{-\beta H + \beta \mu_B N_B}} \over Z}
\end{equation}
 Under the substitution $e^{-\beta H} \rightarrow e^{-itH}$, the partition function becomes the time
 evolution operator of QCD.  Therefore, if we change $t \rightarrow it$,and redefine zeroth
 components of fields by  appropriate factors of i, and introduce Euclidean gamma matrices with anti-commutation relations
 \begin{equation}
   \{ \gamma^\mu, \gamma^\nu \} = -2 \delta^{\mu \nu}
 \end{equation}
  then for QCD, the partition function has the path integral representation 
  \begin{equation}
  Z = \int~[dA] [d\overline \psi ] [d\psi] exp\left\{ -\int_0^\beta~ d^4x~\left( {1 \over 4 }F^2 
  +\overline \psi \left[ {1 \over i} \gamma \cdot D + m+ i \mu_Q \gamma^0 \psi \right]\right) \right\}
  \end{equation}
Here the fermion field is a quark field so that the baryon number chemical potential is
\begin{equation}
  \mu_Q = {1 \over N_c} \mu_B
\end{equation}
This path integral is in Euclidean space and is computable using Monte Carlo methods when
the quark chemical potential vanishes.  If the quark chemical potential is non-zero, various contributions appear with different sign, and the Monte Carlo integrations are poorly convergent.  Boundary conditions
on the fields must be specified on account of the finite length of the integration in time.  They are periodic for Bosons and anti-periodic for Fermions, and follow from the trace in the definition of the partition function.

A straightforward way to probe the confining properties of the QCD matter is to introduce a heavy
test quark.  If the free energy of the heavy test quark is infinite, then there is confinement,
and if it is finite there is deconfinement.  We shall see below that the free energy of an quark added to the system is
\begin{equation}
  e^{-\beta F_q} = <L>
  \label{L}
\end{equation}
where
 \begin{equation}
    L(\vec{x}) = {1 \over N_c} Tr~ P~ e^{i \int ~dt~ A^0(\vec{x},t)}
 \end{equation}
So confinement means $<L> = 0$ and deconfinement means that $<L>$ is finite.  The path ordered phase integration which defines the line operator $L$ is shown in Fig. \ref{line}.  Such a path ordered phase  is called a Polyakov loop or Wilson line.
\begin{figure}[ht]
        \begin{center}
        \includegraphics[width=0.50\textwidth]{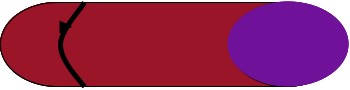}
        \end{center}
        \caption{The contour in the t plane which defines the Polyakov loop.  The space is closed in time because of the periodic boundary conditions imposed by the definition of the partition function.}
\label{line}
\end{figure}

It is possible to prove that the free energy of a heavy static quark added to the system is given by Eqn. \ref{L}
using the effective action for a very heavy quark:
\begin{equation}
 S_{HQ} = \int ~dt ~\overline \psi (\vec{x},t) ~{1 \over i}\gamma^0 D^0~ 
\psi (\vec{x},t).
\end{equation}
The Yang-Mills action is invariant under gauge transformations that are periodic up to an element of the center of the gauge group.  The center of the gauge group is a set of diagonal matrices matrix $Z_p = e^{2\pi i p/N} \overline I$ where $\overline I$ is an identity matrix.
The quark contribution to the action is not invariant, and  $L \rightarrow Z_p L$ under this transformation.  In a theory with only dynamical gluons, the energy of a system of $n$ quarks minus antiquarks is invariant under the center symmetry transformation only if $n$ is an integer multiple of $N$.  Therefore, when the center symmetry is realized, the only states of finite free energy
are baryons plus color singlet mesons.

The realization of the center symmetry, $L \rightarrow Z_p L$
is equivalent to confinement.  This symmetry is like the global rotational symmetry of a spin system, and it may be either realized or broken.   At large separations, the correlation of a line and its adjoint, corresponding to a quark-antiquark pair is
\begin{equation}
        lim_{r \rightarrow \infty} <L(r) L^\dagger (0)> = Ce^{-\kappa r} + <L(0)><L^\dagger (0)>
\end{equation}        
since upon subtracting a mean field term, correlation functions should vanish exponentially.  Since
\begin{equation}
   e^{-\beta F_{q \overline q} (r)} = <L(r) L^\dagger (0)>
\end{equation}
we see that in the confined phase, where $<L> = 0$, the potential is linear, but in the unconfined phase,
where $<L>$ is non-zero, the potential goes to a constant at large separations.

The analogy with a spin system is useful.  For the spin system corresponding to QCD
without dynamical quarks,
the partition function  can be written as
\begin{equation}
 Z = \int~ [dA]~e^{- {1\over g^2} S[A]}
\end{equation}
The effective temperature of the spin system associated with the gluon fields is $T_{eff} \sim g^2$.
By asymptotic freedom of the strong interactions, as real temperature gets larger, the effective temperature gets smaller.  So at large real temperature (small effective temperature) we expect an ordered system, where the $Z_N$ symmetry is broken, and there is deconfinement.  For small real temperature corresponding to large effective temperature, there is disorder or confinement.

The presence of dynamical fermions breaks the $Z_N$ symmetry.  This is analogous to placing a spin system in an external magnetic field.  There is no longer any symmetry associated with confinement, and the phase transition can disappear.  This is what is believed to happen in QCD for physical masses of quarks.  What was a first order phase transition for the theory in the absence of quarks becomes  a continuous change in the properties of the matter for the theory with quarks.

\begin{figure}[ht]
        \begin{center}
        \includegraphics[width=0.60\textwidth]{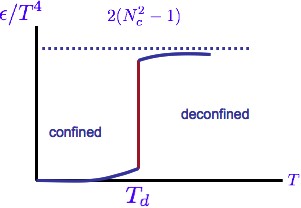}
        \end{center}
        \caption{The energy density scaled by $T^4$ for QCD in the absence of dynamical quarks.}
\label{et4}
\end{figure}

Another way to think about the confinement-decofinement transition is a change in the number of degrees of freedom.  At low temperatures, there are light meson degrees of freedom.  Since these
are confined, the number of degrees of freedom is of order one in the number of colors.  In the unconfined world, there are $2(N_c^2-1)$ gluons, and $4N_cN_f$ fermions where $N_f$ is the number of light mass fermion families.  The energy density scaled by $T^4$ is a dimensionless number and directly proportional to the number of degrees of freedom.  We expect it to have the property shown in Fig. \ref{et4} for pure QCD in the absence of quarks.  The discontinuity at the deconfinement temperature, $T_d$ is the latent heat of the phase transition.

\begin{figure}[ht]
        \begin{center}
        \includegraphics[width=0.60\textwidth]{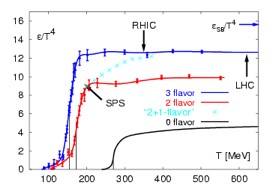}
        \end{center}
        \caption{The energy density scaled by $T^4$ measured in QCD from lattice Monte Carlo simulation.  Here there are quarks with realistic masses.}
\label{et4lat}
\end{figure}

The energy density can be computed using lattice Monte Carlo methods.  The result of such computation is shown in Fig. \ref{et4lat}.  The discontinuity present for the theory with no quarks becomes a rapid cross over when dynamical quarks are present.

The large $N_c$ limit gives some insight into the properties of high temperature 
matter\cite{'tHooft:1973jz}-\cite{Thorn:1980iv}. As $N_c \rightarrow \infty$, the energy density itself is an order parameter for the decofinement phase transition.  Viewed from the hadronic world, there is an amount of energy density $\sim N_c^2$ which must be inserted 
to surpass the transition temperature.  At infinite $N_c$ this cannot happen, as this involves an infinite amount of energy.  There is a Hagedorn limiting temperature, which for finite $N_c$ would have been the deconfinement temperature.

The Hagedorn limiting temperature can be understood from the viewpoint of the hadronic world as arising from an exponentially growing density of states.  In a few paragraphs, we will argue that mesons and glueballs are very weakly interacting in the limit of large $N_c$.  Therefore, the partition function is
\begin{equation}
  Z = \int~ dm~\rho (m) e^{-m/T}
 \end{equation}
 Taking $\rho(m) \sim m^\alpha e^{\kappa m}$, so that 
 \begin{equation}
   <m> \sim {1 \over {1/T-\kappa}}
 \end{equation} 
diverges when $T \rightarrow 1/\kappa$

\subsection{A Brief Review of the Large $N_c$ Limit}

\begin{figure}[htbp]
\begin{center}
\begin{tabular} {l l l}
\includegraphics[width=0.50\textwidth] {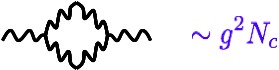}  & &
\includegraphics[width=0.46\textwidth] {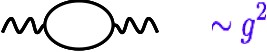} \\
& & \\
a & & b \\
\end{tabular}
\end{center}
\caption{a:  The gluon loop contribution to the heavy quark potential.  b:  The quark loop contribution to the potential}
\label{loop}
\end{figure}

The large $N_c$ limit for an interacting theory takes $N_c \rightarrow \infty $ with the 't Hooft coupling
$g^2_{'t Hooft} = g^2 N_c$ finite.  This approximation has the advantage that the interactions among quarks and gluons simplify.  For example, at finite temperature, the disappearance of confinement
is associated with Debye screening by gluon loops, as shown in Fig. \ref{loop}a.  This diagram generates a screening mass of order $M^2_{screening} \sim g^2_{'t Hooft} T^2$.   On the other hand the quark loop contribution is smaller by a power of $N_c$ and vanishes in the large $N_c$ limit.

\begin{figure}[htbp]
\begin{center}

a  \includegraphics[width=0.75\textwidth ] {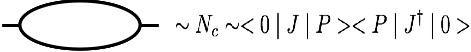}  \\
 ~ \\
~~b ~~~ \includegraphics[width=0.75\textwidth ] {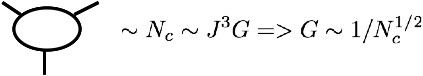} \\

\end{center}
\caption{a:  The quark loop corresponding to a current-current interaction.  b: A quark loop corresponding to a three current interaction.}
\label{int}
\end{figure}

To understand interactions, consider Fig. \ref{int}a.  This corresponds to a mesonic current-current interaction through quarks.  In powers of $N_c$, it is of order $N_c$.  Gluon interactions will not change this overall factor.  The three current interaction is also of order $N_c$ as shown in Fig. \ref{int}b.  The three meson vertex, $G$ which remains after amputating the external lines, is therefore of order $1/\sqrt{N_c}$.  A similar argument shows that the four meson interaction is of order $1/N_c$.
Using the same arguments, one can show that the 3 glueball vertex is of order $1/N_c$ and the four glueball interaction of order
$1/N_c^2$.

These arguments show that QCD at large $N_c$ becomes a theory of non-interacting mesons and glueballs.  There are an infinite number of such states because excitations can never decay.  In fact, the spectrum of mesons seen in nature does look to a fair approximation like non-interacting particles.  Widths of resonances are typically of order $200~ MeV$, for resonances with masses up to several $GeV$.
\subsection{Mass Generation and Chiral Symmetry Breaking}

QCD in the limit of zero quark masses has a $U(1) \times SU_L(2) \times SU_R(2)$ symmetry.  (The $U_5(1)$ symmetry is explicitly broken due to the axial anomaly.)  Since the pion field, $\overline \psi \tau^a \gamma_5 \psi$ is generated by an $SU_{L-R}(2)$ transformation of the sigma field, $\overline \psi \psi$, the energy (or potential) in the space of the pion-sigma field is degenerate under this transformation.
In nature, pions have anomalously low masses.  This is believed to be a consequence of chiral symmetry breaking, where the $\sigma $ field acquires an expectation value, and the pion fields are Goldstone bosons associated with the degeneracy of the potential under the chiral rotations.

\begin{figure}[ht]
        \begin{center}
        \includegraphics[width=0.40\textwidth]{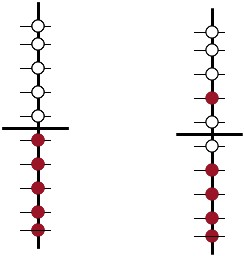}
        \end{center}
        \caption{The energy levels of the Dirac equation.  Unfilled states are open circles and filled states are solid circles.  For the free Dirac equation, negative energy states are filled and positive energy states are unoccupied, as shown on the left hand side.  A mesonic excitation corresponding to a particle hole pair is shown on the right hand side.}
\label{hole}
\end{figure}

Such symmetry breaking can occur if the energy of a particle-antiparticle pair is less than zero, as shown in Fig. \ref{hole}.  On the left of this figure is the naive vacuum where the negative energy states associated with quark are filled.  The right hand side of the figure corresponds to a particle hole excitation, corresponding to a sigma meson.  Remember that a hole in the negative energy sea corresponds to an antiparticle with the opposite momentum and energy. If the $\sigma$ meson excitation has negative energy, the system is unstable with respect to forming a condensate of these mesons.

At sufficiently high temperature, the chiral condensate might melt.  Indeed this occurs\cite{Karsch:2001cy} .For QCD,
the chiral and deconfinement phase transition occur at the same temperature.  At a temperature of about $170 - 200~ MeV$, both the linear potential disappears and chiral symmetry is restored.  It is difficult to make a precise statement about the indentification of the chiral and deconfinement phase transitions,
since as argued above, for QCD with quarks, there is not a real phase transition associated with deconfinement\cite{Bazavov:2009zn}-\cite{Borsanyi:2010cj}.
Also, when quarks have finite masses, as they do in nature, chiral symmetry is not an exact symmetry, and there need be no strict phase transition associated with its restoration.  Nevertheless, the cross over is quite rapid, and there are rapid changes in the both the potential and the sigma condensate $<\overline \psi \psi >$ at temperatures which are in a narrow range.

\section{Lecture III: Matter at High Baryon Number Density: Quarkyonic Matter}

I now turn to a discussion of the phase diagram of QCD at finite baryon number density.

In the large $N_c$ limit of QCD, the nucleon mass is of order $N_c$\cite{'tHooft:1973jz}-\cite{Witten:1979kh}.  This means that in the confined phase of hadronic matter, for baryon chemical potential $\mu_B \le M_N$, the baryon number density is
essentially zero:
\begin{equation}
   <N_B> \sim e^{(\mu_B-M_N)/T} \sim e^{-N_c}
\end{equation}   
For temperatures above the de-confinement phase transition the baryon number is non-zero since there the baryon number density is controlled by $e^{-M_q/T} \sim 1$, and quark masses are independent of
$N_c$.  For sufficiently large chemical potential the baryon number density can be nonzero also. The Hadronic Matter phase of QCD is characterized in large $N_c$ by zero baryon number density, but at higher density there is a new phase.
 \begin{figure}
\centering
\includegraphics[scale=0.30]{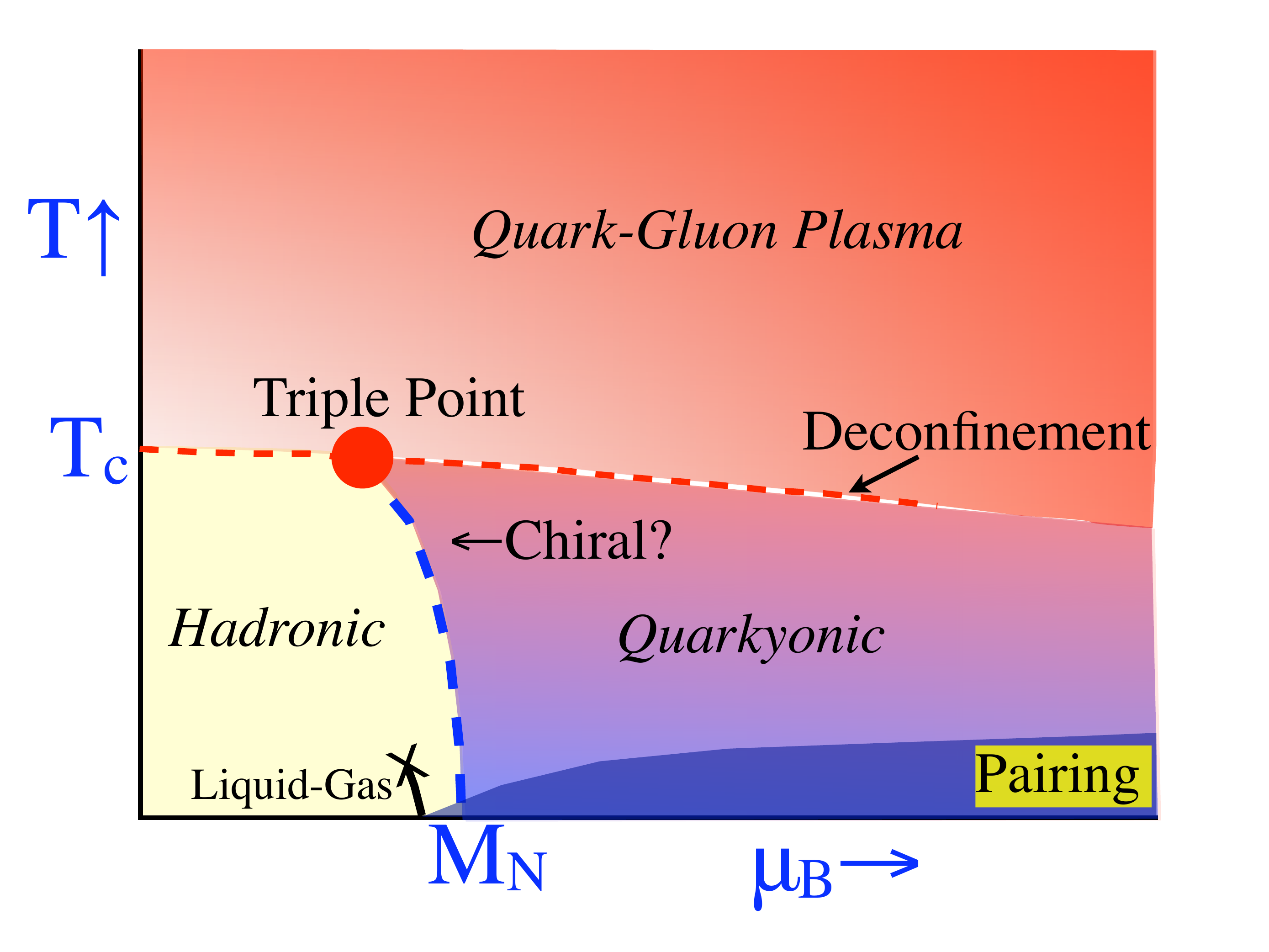}	       
\caption[]{The revised phase diagram of QCD}
\label{phasediagram}
\end{figure}

In the large $N_c$ limit, fermion loops are suppressed by a factor of $1/N_c$.  Therefore the contribution to Debye screening from quarks cannot affect the quark potential until
\begin{equation}
  M_{Debye}^2 \sim \alpha_{t'Hooft}~ \mu_{quark}^2/N_c \sim \Lambda^2_{QCD}
\end{equation}
Here the quark chemical potential is $\mu_B = N_c \mu_{quark}$.  The relationship involving the Debye mass means there is a region parametrically large  chemical potential $M_N \le \mu_B \le \sqrt{N_c}M_N$ where matter is confined, and has finite baryon number.  This matter is different than either the Hadronic Matter or the De-Confined Phases.  It is called Quarkyonic because it exists at densities parametrically large compared to the QCD scale, where quark degrees of freedom are important,
but it is also confined so the degrees of freedom may be thought of also as those of confined 
baryons\cite{McLerran:2007qj}-\cite{Hidaka:2008yy}.

The width of the transition region between the Hadronic phase and the Quarkyonic phase is estimated
by requiring that the baryon number density become of order $N_B/V \sim k_{Fermi}^3 \sim \Lambda_{QCD}^3$.  Recall that the baryon chemical potential is $\mu_B \sim M_N + k_f^2/2M_N$ for small $k_F$, so that the width of the transition in $\mu_B$ is very narrow, of order $1/N_c$. This is $\delta \mu_{qaurk} \sim  1/N_c^2$ when expressed in terms of $\mu_{quark}$ which is the finite variable in the large $N_c$ limit.
 \begin{figure}
\centering
\includegraphics[scale=0.60]{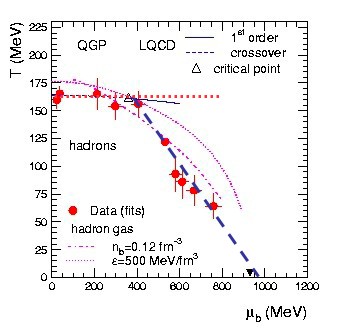}	       
\caption[]{Chemical potentials and temperatures at decoupling.}
\label{line}
\end{figure}

The transition from Hadronic Matter to that of the Quark Gluon Plasma may be thought of as a change in the number of degrees of freedom of matter.  Hadronic Matter at low temperatures has  3 pion degrees of freedom.  The quark gluon plasma has of order $2(N_c^2-1)$ degrees of freedom corresponding to gluons and $4 N_c$ degrees of freedom for each light mass quark.  The change in degrees of freedom is of order $N_c^2$ in the large $N_c$ limit.  At very high baryon number densities, the quarks in the Fermi sea interact at short distances, and although strictly speaking are confined, behave like free quarks.  The number of degrees of freedom is therefore of order $N_c$.  Each phase has different numbers of degrees of freedom, and is presumably separated from the other by a rapid crossover.
 \begin{figure}
\centering
\includegraphics[scale=0.60]{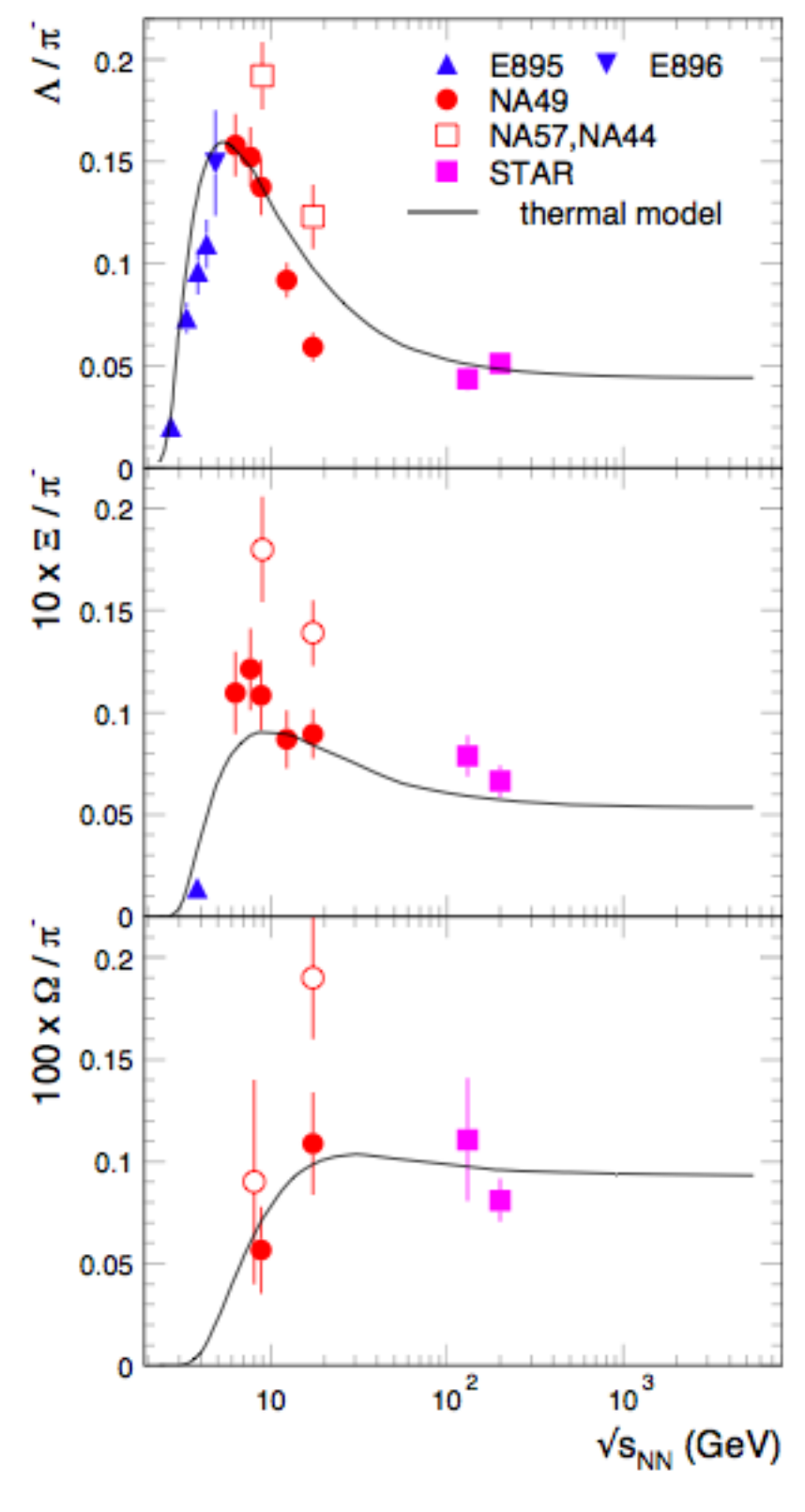}	       
\caption[]{Ratios of abundances of various particles .}
\label{horn}
\end{figure}
Quarkyonic matter is confined and therefore thermal excitations such as mesons, glueballs, and Fermi surface excitations must be thought of as confined.  The quarks in the Fermi sea are effectively weakly interacting since their interactions take place at short distances.  So in some sense, the matter is ``de-confined" quarks in the Fermi sea with confined glueball, mesons and Fermi surface excitations\cite{Castorina:2010gy}.

In Hadronic Matter, chiral symmetry is broken and in Deconfined Matter it is broken.  In Quarkyonic Matter chiral symmetry is broken by the formation of charge density waves from binding of quark and quark hole excitations near the Fermi surface\cite{Deryagin:1992rw}.  In order that the quark hole have small relative momentum to the quark, the quark hole must have momentum opposite to that of the quark.  This means the quark-quark hole excitation has total net momentum, and therefore the finite wavelength of the corresponding bound state leads to a breaking of translational invariance.  The chiral condensate turns out to be a chiral spiral where the chiral condensate rotates between different Goldstone boson
as one moves through the condensate\cite{Kojo:2009ha}.  Such condensation may lead to novel crystalline structures\cite{tsvelik}.

A figure of the hypothetical phase diagram of QCD is shown in Fig. \ref{phasediagram} for $N_c = 3$.  Also shown is the weak liquid-gas phase transition, and the phase associated with color superconductivity.  Although the color superconducting phase cannot coexist with quarkyonic matter in infinite $N_c$, for finite $N_c$ there is such possibility.  The lines on this phase diagram might correspond to true phase transitions or rapid cross overs.  The confinement-deconfinement transition is known to be a cross over.  In the FPP-NJL model\cite{Fukushima:2003fw}-\cite{Pisarski:2000eq}, the Hadronic-Quarkyonic transition is first order\cite{McLerran:2008ua},  but nothing is known from lattice computations.  If as we conjecture, there is region where chiral symmetry is broken by translationally non-invariant modes, then this region must be surrounded by a line of phase transitions.  I call this region Happy Island becuase it is an island of matter in the $\mu_B-T$ plane.

A remarkable feature of this plot is the triple point where the Hadronic Matter, Deconfined Matter and
Quarkyonic Matter all meet\cite{Andronic:2009gj}.  This triple point is reminiscent of the triple point for the liquid, gas and vapor phases of water. 

Since we expect a rapid change in the number of degrees of freedom across the transitions between
these forms of matter, an expanding system crossing such a transition will undergo much dilution would undergo much dilution at a fixed value of temperature or baryon chemical potential\cite{BraunMunzinger:1994xr}-\cite{Heinz:1999kb}.  One might expect in heavy ions to see decoupling of particle number changing processes at this transition, and the abundances of produced particles will be characteristic of the transition.  In Fig. \ref{line}

In the Fig. \ref{line}, the expectations for the confinement-deconfinement transition are shown with the dotted red line.  It is roughly constant with the baryon chemical potential, and the constant value of temperature is taken from lattice estimates.  The dark dashed curve represents $\mu_B -T = cons \times M_N$, corresponding to a simple model for the Quarrkyonic transition.  Such a very simple description does remarkably well.  

A triple point is suggested at a baryon chemical potential near 400 MeV, and temperature near 160 MeV.  This corresponds to a center of mass energy for Pb-Pb collisions of 9-10 GeV.  This is near where there are anomalies in the abundances of rations of particles\cite{Gazdzicki:1998vd}, as shown in Fig. \ref{horn}.
Shown are fits using statistical models of abundances of particles using chemical potentials and temperature extracted from experimental data.  The sharp peak reflects the change in behavior as one proceeds along the dashed line of Fig. \ref{line} corresponding to the Quarkyonic transition and joins to the dotted red line 
of the deconfinement transition

\begin{figure}
\centering
\includegraphics[scale=0.60]{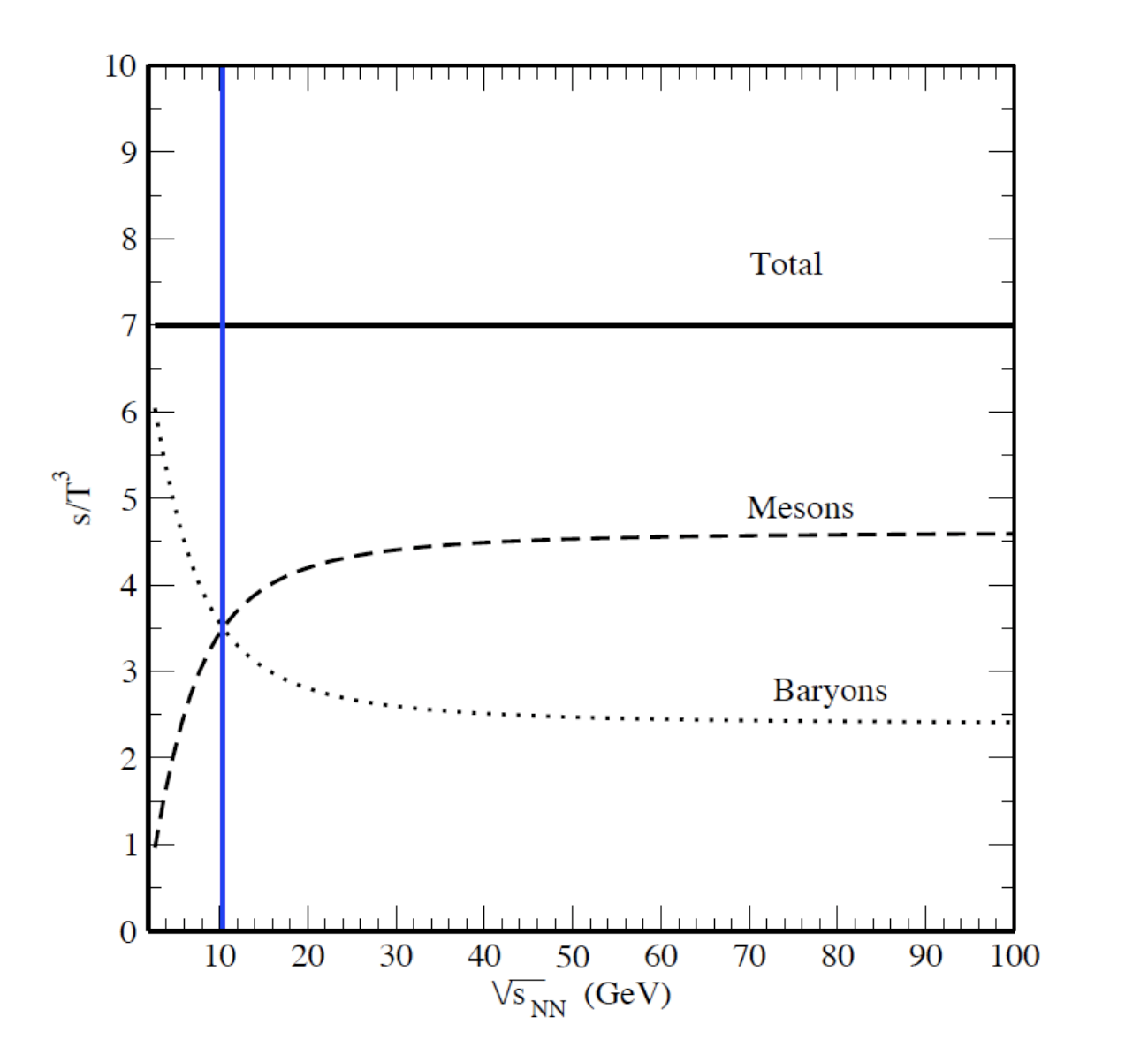}	       
\caption[]{Energy density stored in baryons compared to that stored in mesons.}
\label{baryons}
\end{figure}

It is remarkable that the value of beam energy where this occurs corresponds to the hypothetical triple point of Fig. \ref{line}, and that this is the density where the energy density stored in baryons becomes equal to that stored in mesons, Fig. \ref{baryons},

\section*{Acknowledgments}

I gratefully acknowledge the organizers of the 50'th Crakow School of Theoretical Physics, in particular,
Michal Praszalowicz, for making this wonderful and extraordinary meeting.
The research of  L. McLerran is supported under DOE Contract No. DE-AC02-98CH10886.

\end{document}